\documentclass[sigconf]{acmart}
\usepackage{booktabs} 
\usepackage{color, colortbl}
\definecolor{Gray}{gray}{0.92}
\usepackage{framed}
\definecolor{Gray}{gray}{0.93}

\begin{document}
\settopmatter{printacmref=false}
\renewcommand\footnotetextcopyrightpermission[1]{} 
\pagestyle{plain}

\title{Find, Understand, and Extend Development \\ Screencasts on YouTube}

\author{Mathias Ellmann}
\affiliation{
  \institution{University of Hamburg}}

\email{ellmann@informatik.uni-hamburg.de}

\author{Alexander Oeser}
\affiliation{
  \institution{University of Hamburg}}

\email{9oeser@informatik.uni-hamburg.de}

\author{Walid Maalej}
\affiliation{
  \institution{University of Hamburg}}

\email{maalej@informatik.uni-hamburg.de}

\author{Davide Fucci}
\affiliation{
  \institution{University of Hamburg}}

\email{fucci@informatik.uni-hamburg.de}

\begin{abstract}
A software development screencast is a video that captures the screen of a developer working on a particular task while explaining its implementation details. 
Due to the increased popularity of software development screencasts (e.g., available on YouTube), we study how and to what extent they can be used as additional source of knowledge to answer developer's questions about, for example, the use of a specific API. 
We first differentiate between development and other types of screencasts using video frame analysis. 
By using the Cosine algorithm, developers can expect ten development screencasts in the top 20 out of 100 different YouTube videos. 
We then extracted popular development topics on which screencasts are reporting on YouTube: 
database operations, system set-up, plug-in development, game development, and testing. 

Besides, we found six recurring tasks performed in development screencasts, such as object usage and UI operations.
 
Finally, we conducted a similarity analysis by considering only the spoken words (i.e., the screencast transcripts but not the text that might appear in a scene) to link API documents, such as the Javadoc, to the appropriate screencasts.
By using Cosine similarity, we identified 38 relevant documents in the top 20 out of 9455 API documents. 

\end{abstract}


\maketitle

\section{Introduction}

Software development is a knowledge-intensive work \cite{Maalej:TOSEM:2014, Sillito:TSE:2008, Ko:TSE:2006, Fritz:ICSE:2010} in which developers spend a substantial amount of their time looking for information \cite{Ko:TSE:2006}---e.g., how to fix a bug or how to use an API. 
They access and share knowledge through various media and sources, including API documentation \cite{Maalej:TSE:2013}, Q\&A sites, wikis, or tutorials \cite{treude2016augmenting, WhatIsSc21:online, Maalej:TOSEM:2014}. 
Regardless of how rich or popular a single knowledge source might be, it barely satisfies all the information needs of a specific developer within a certain context \cite{Maalej:ASE:2009, Fritz:ICSE:2010, Maalej:TOSEM:2014}. 

Nowadays, there is a growing trend to use videos instead of text to capture and share knowledge \cite{lethbridge2003software}. 
Video content, from movies to webinars and screencasts, accounts for more than half of the internet traffic\footnote{\url{http://www.cisco.com/c/en/us/solutions/collateral/service-provider/visual-networking-index-vni/complete-white-paper-c11-481360.html}}. 
Software developers are concerned with this trend as they are using more and more video resources in their job \cite{macleod2015code, tiarks2014does}.
In particular, development screencasts are getting popular among technical bloggers\footnote{http://www.virtuouscode.com/a-list-of-programming-screencast-series/ \\
https://www.rubytapas.com/new-list-programming-screencast-series/} and on general purpose video-sharing platforms such as YouTube.

A screencast is a \textit{``digital movie in which the setting is partly or wholly a computer screen, and in which audio narration describes the on-screen action''} \cite{WhatIsSc21:online}. 
In particular, a development screencast is created by a developer to describe and visualize a certain development task \cite{macleod2015code}. 
Screencasts are more comprehensive than plain text since they capture, in the form of video and audio, human interaction\cite{laptev2008learning}---e.g., following the instruction of a developer. 

YouTube\footnote{\url{https://support.google.com/youtube/answer/4594615?hl=en}} does not yet offer the possibility to explicitly search for a development screencast that explains how to accomplish a specific development task \cite{Maalej:JSS:2016, treude2016augmenting} in a certain development context \cite{Maalej:CSD:2015, Maalej:JSS:2016}. 

Moreover, there is a lack of understanding about the different types of videos—i.e., development screencast \cite{macleod2015code} cannot be distinguished from other types of videos.

\begin{figure}[!t]
\centering
\includegraphics[width=0.45\textwidth]{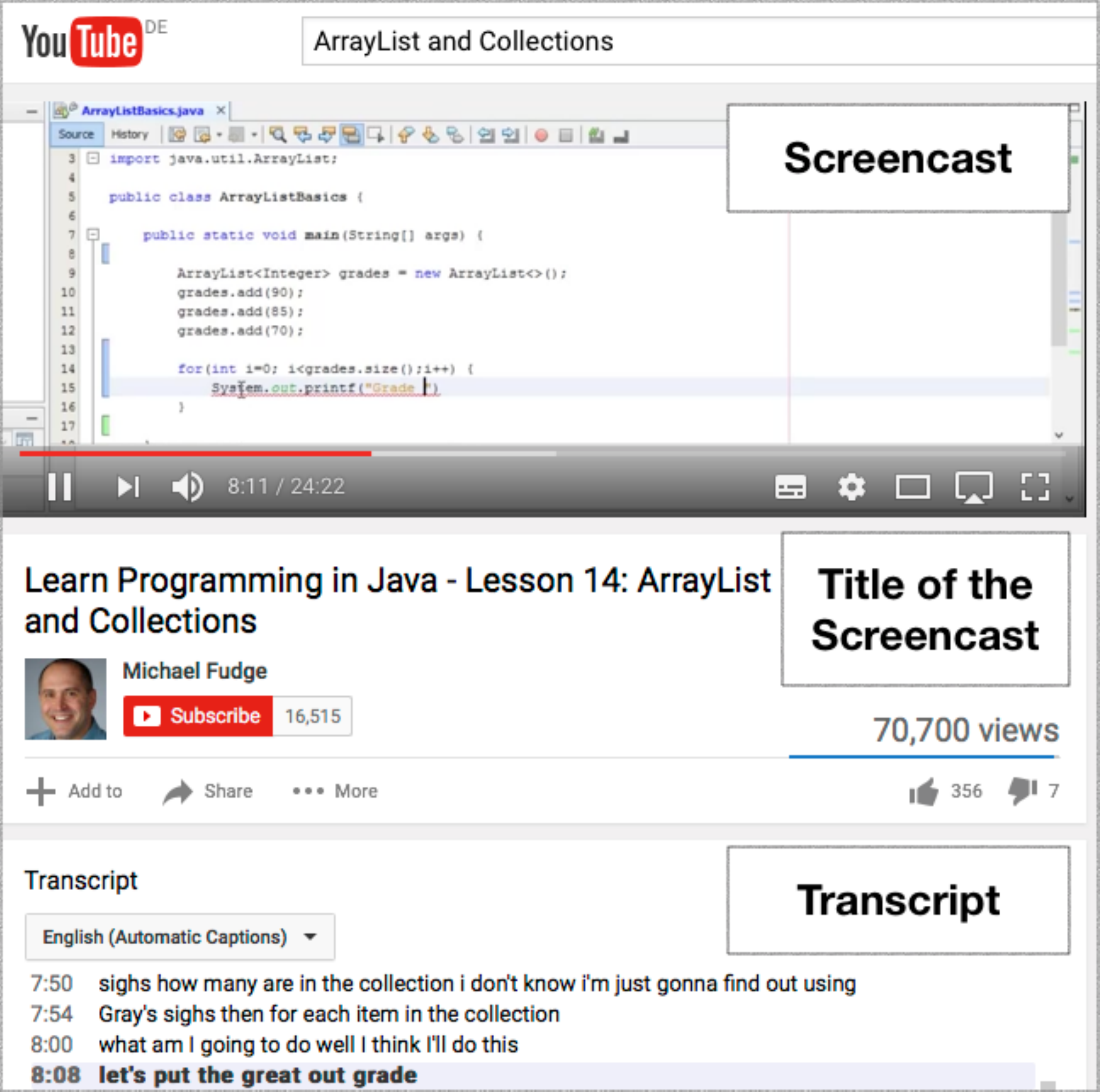}
\caption{Example of a development screencast on YouTube. It contains a video (screencast), a title describing the software development task, and a transcript.}
\label{fig:Screencast}
\end{figure}

\begin{figure}[t!]
\centering
   \includegraphics[width=0.45\textwidth]{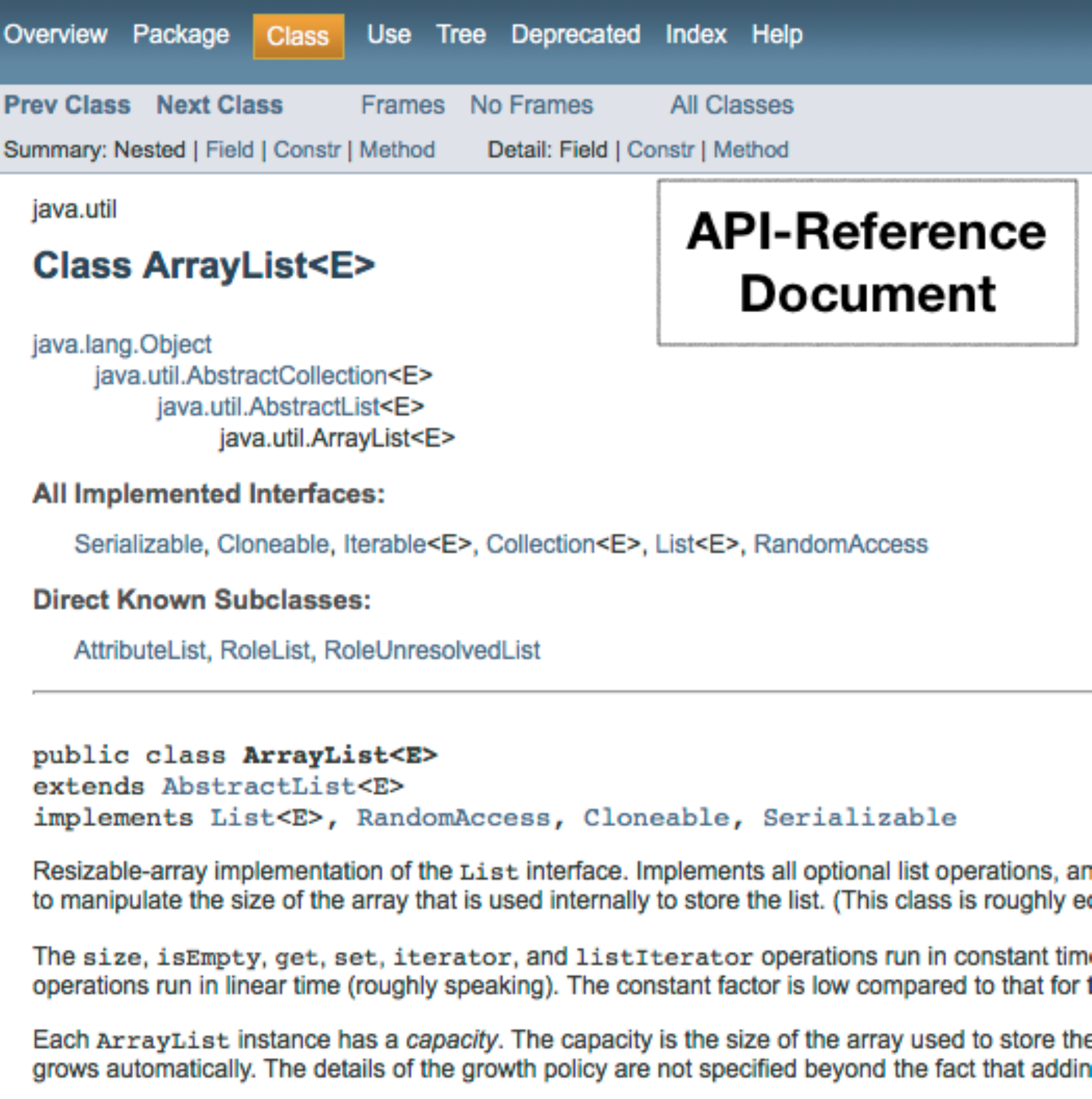}
 \caption{Example of an API reference document. It contains class and method definitions as well as the descriptions of it.}
 \label{fig:API_ref}
\end{figure}

In a development screencast, a software developer performs a task which can be assigned to a topic and to a specific context \cite{Maalej:CSD:2015, macleod2015code}, such as an IDE, a web browser or a virtual machine. 
There are recurring tasks performed in several development screencasts and in different development contexts that require the consultation of API documents\cite{treude2015tasknav, Maalej:TSE:2013}. 
The text transcript of the screencast audio contains searchable and indexable technical terms that can refer to other artifacts, such as an API or  a tool. 
For example, the screencast presented in Figure \ref{fig:Screencast} can be extended with an API document---as shown in Figure \ref{fig:API_ref}---since it contains references to classes, methods and other units.  

Based on the mentioned observations, we will tackle the following research questions in this paper:
\begin{enumerate}
\item \textit{RQ1}: Is it possible to distinctively identify a developer screencasts from other video types based on a frame analysis? 
\item \textit{RQ2}: Which development tasks are performed in software development screencasts?
\item \textit{RQ3}: Can a development screencast be extended with relevant API reference documents by considering only the spoken words? 
\end{enumerate}

In particular, in Section \ref{sec:frame_analysis}, we evaluate different algorithms (Jaccard, Cosine \& LSI) and their performance in identifying development screencasts by simply considering video frames. 
In Section \ref{sec:topic_analysis}, we use the visualization techniques introduced by Sievert et al. \cite{sievert2014ldavis} and Chuang et al. \cite{chuang2012termite} to identify the software development topics and the recurring development tasks present in  development screencasts. 
In Section \ref{sec:similarity_analysis}, we analyse the similarity between a task performed in the screencast and the relevant API documents using the TaskNav tool \cite{treude2015tasknav}. 
Section \ref{sec:discussion} discusses the results, while Section \ref{sec:conclusion} concludes the paper and describes future work.

\label{sec:frame_analysis}

A development screencast is a special type of video which cannot be directly searched on YouTube due to the lack of a pre-defined category. 
Nonetheless, a development screencast is characterized by the small number of scenes, length, and the specific actions (e.g., inputting text) performed by a developer \cite{macleod2015code}. 
Moreover, in their screencasts, developers use several tools (e.g., and IDE or code editor, a terminal, a browser) to perform a development task. 
In this section, we present how we used the information available on the video frames to distinguish a development screencast from other types of videos.

We sampled a set of frames (i.e., a rectangular raster of pixels) from different videos and compared their stability and variation.
We define the similarity between two frames as $Sim(f_1,f_2)$. 
The frame similarity of a video is calculated by \( \sum \limits_{1}^{n} \frac{ Sim(f_n,f_{n+1})}{n} \) with \textbf{n} as the number of analyzed frames.
Frame $f_{n+1}$ is the direct successor of frame $f_n$ and $f_n \neq f_{n+1}$. 
For each video, we sampled a frame every 10 seconds.

We randomly selected 100 YouTube videos associated to one of the following types:
\begin{itemize}
\item \textbf{Development screencast (n=20)}: 
videos showing  software development activities  
in several programming languages, such as PHP, Python, Java, SQL and C\#. 
 Different tools (e.g., an IDE, code editor, or simple a web browser) are used to perform a task.  
 \item \textbf{Non-development screencasts (n=20)}: 
 videos showing the desktop of a user solving problems unrelated to software development, including mathematical problems, game tutorials, or software utilization.
 \item \textbf{Non-development, non-screencast (n=20)}:	 videos showing how to perform a task not related to software development (e.g., learn Spanish, or how-to change a phone screen) in which a computer screen is not recorded.
 \item \textbf{Others (n=40)}\footnote{Provided from the owner of http://randomyoutube.net}: 
 videos in none of the above categories (e.g., a music video). This set contains 40 videos because most of them had a short length (2-3 minutes).
\end{itemize}

The sample contains approximately 2000 frames for each video type.
Every frame contains a particular number of color information that changes in the different scenes throughout the developer screencast---for example, when using an IDE, a web browser or a terminal.

\begin{figure}[h]
\centering
   \includegraphics[width=0.49\textwidth]{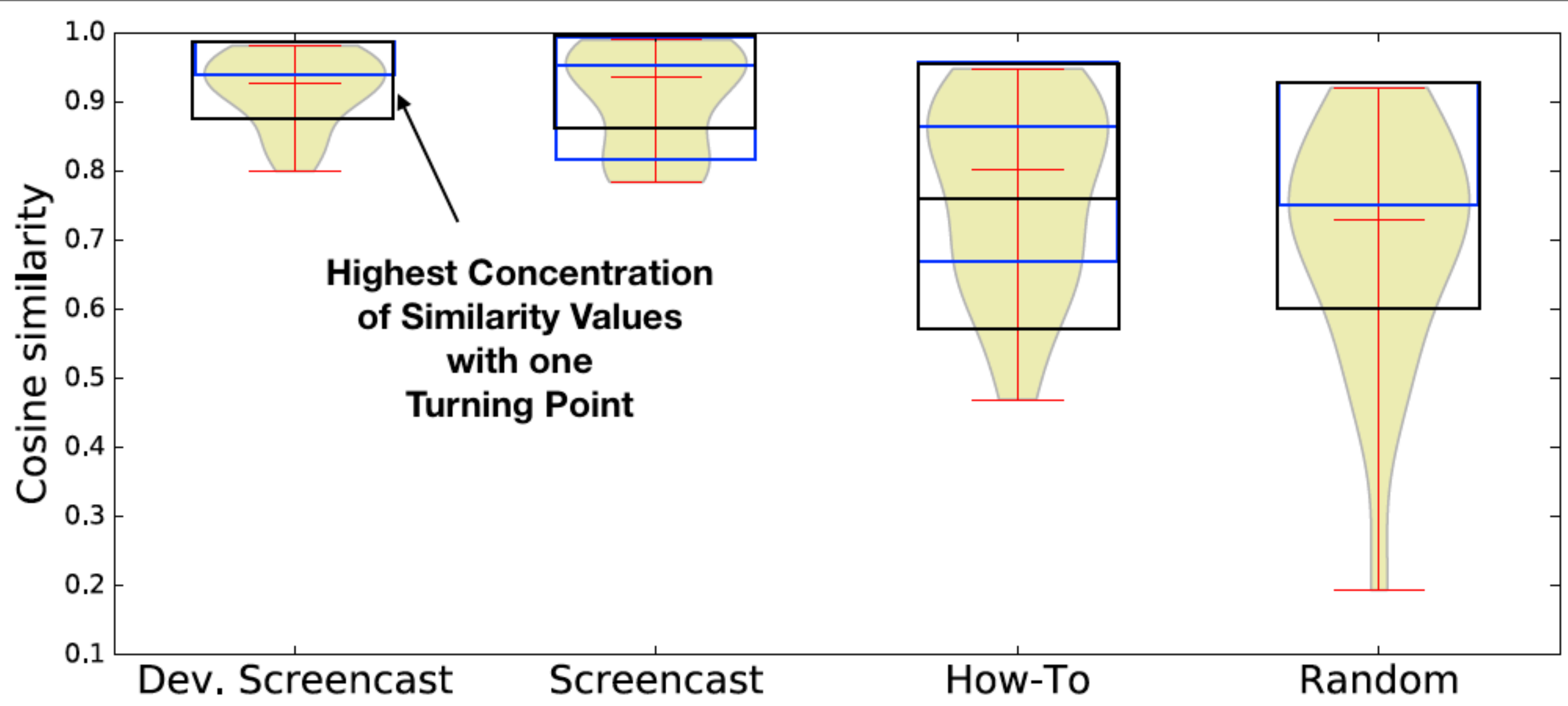}
 \caption{Frame similarity of development screencasts compared to other video types (using cosine similarity values).}
 \label{fig:frame_similarity}
\end{figure}

The similarity between two frames was calculated using the \textit{Jaccard} coefficient, \textit{Cosine} similarity, and \textit{LSI}.  
Each color information per pixel is considered a bag of words \cite{wang2008spatial}.
The \textit{Jaccard} coefficient is used to measure the similarity between two sets of data. 
The cardinality of the intersection is divided by the cardinality of union of the sets \cite{huang2008similarity}. 
The similarity value of the \textit{Jaccard} coefficient ranges between 0 and 1. 
If the documents \(\vec{t_a}\) and \(\vec{t_b}\) contain the same set of objects, the coefficient is one (or zero in case the documents do not have objects in common). 
The similarity between the two documents  \(\vec{t_a}\) and \(\vec{t_b}\) is \[ SIM_J(\vec{t_a},\vec{t_b}) = \frac{|\vec{t_a} \cdot \vec{t_b}|}{|\vec{t_a}|^2 + |\vec{t_b}|^2 - \vec{t_a} \cdot \vec{t_b}}\]
The \textit{Cosine} approach is commonly used for a similarity comparison of documents \cite{ahasanuzzaman2016mining},\cite{conrad2003online},\cite{park2002analysis}.
Documents are converted into term vectors to compute their \textit{Cosine} similarity, which is quantified as the angle between these vectors and ranges between 0 and 1\cite{huang2008similarity}. 
Finally, the LSI ranges between -1 and +1; it uses term frequencies for dimensionality reduction, followed by a singular value decomposition (SVD)\cite{mihalcea2006corpus}. 
We use the \textit{Cosine} and \textit{LSI} algorithms to evaluate the frequency of scene switches in a video. 
The \textit{Jaccard} algorithm is more sensitive than \textit{Cosine} and \textit{LSI} as the latter two only recognize a low number of scene switches and moving objects (mouse, keyboard, etc.) used in the development screencasts.

The analysis of 2127 frames from 20 development screencasts shows that the values of \textit{Cosine} and \textit{LSI} are close to 1.0, indicating that, in a development screencasts, there is only a small number of scene switches.
The Jaccard similarity has an average value of 0.768, showing that small objects are moved.  
We analyzed the similarity distributions of the four sets of videos using each algorithm.

The highest concentration of similar values for the development screencasts can be calculated using the Cosine algorithm (see Figure \ref{fig:frame_similarity}). 
For the Jaccard algorithm, the characteristics of the distribution varies a lot,  making it difficult to identify a developer screencast from other types of video.
The LSI algorithm has similar distributions. 
On the other hand, the Cosine algorithm shows a higher concentration, particularly for Developer Screencasts. 
Thus, it is  better suited to distinguish developer screencasts from other video types.

We identified 20 development screencast within other video types (n = 100) by using the Cosine algorithm.
We calculated the frame similarity of every video type and ranked the videos based on their Cosine similarity (in descending order). 
All developer screencasts are correctly predicted until the first 45 recommendations due to the high concentration of similarity values for development screencasts with respect to other video types. 
Within a list of 20 videos, we could identify 55\% of the development screencasts. 
In other words, developers can expect to correctly identify over ten development screencast in a list of 20 different YouTube videos with the support of the Cosine algorithm. ($precision$ = 0.028, $recall$ = 0.55, and $F1$ = 0.052 \cite{robillard2014recommendation}). 

To answer RQ1, we found that the Cosine algorithm is better suited to distinguish development screencast from other types of video due to its capability of better concentrating similarity values.
\begin{framed}
\begin{itemize}
\item Development screencasts are different from other types of videos. Development screencasts seem to be more static---i.e., they have less scenes and objects.
\item The Cosine algorithm is the best, among the studied algorithms, at identifying a development screencast from other video types (highest concentration of similarity values). 
\item All development screencasts could be identified within the first third of the retrieved items.
\end{itemize}
\end{framed}

\section{Topics of Software Development Screencasts}
\label{sec:topic_analysis} 
In this section, we analyze the software development topics of the task performed during a development screencast. 
To this end, we analyzed the title of the screencast as well as its audio transcripts and assigned the task performed on screen to different software development topics, such as implementation or system set-up.
\begin{table}[h] 
\caption{Topics of and within Java screencasts.}
\centering
    \begin{center}
        \resizebox{\columnwidth}{!}
                	{
          	  \begin{tabular}{l|l }
          	    \toprule         
   Topic label & Most relevant terms \\
    \midrule
             \multicolumn{2}{l}{6 Topics of development screencasts (tasks performed in SD according to their titles)} \\
          \midrule
           \rowcolor{Gray}
     database operation with Java &  netbean, database, create, mysql  \\
     database operation with Android & class, table, key, android \\ 
     
       \rowcolor{Gray}
    system set-up & run, make, Window, JDK \\ 
    plug-in development & connect, jframe, constructor, jbutton  \\
           \rowcolor{Gray}
    game development      & game, develop, object, implement \\    
     testing & selenium, use, program, file, write, learn \\
    	
    	\midrule
 
    	  \multicolumn{2}{l}{6 Topics within development screencasts (repeatable tasks performed in SD according to the transcripts)} \\        	     	 
    	 \midrule
    	    \rowcolor{Gray}
    	    
   API usage (Object/Classes)  & use, create, class, code, method, click, type   \\
    	           
    	Files        & file, create, call, time, program \\
    	         \rowcolor{Gray}
    	     Lists & list, move, get, create \\

    	         \rowcolor{Gray}
    	         UI operations   & box, file, slider, inputs  \\
    	        Methods &  property, get, input, statement  \\ 
    	         \rowcolor{Gray}
    	       System operations  & program, time, system, get  \\

  	     \bottomrule   
   	  \end{tabular}
   	 }
   	    \end{center}
      \label{table:topic_screencasts}
\end{table}

Due to its popularity, we focused on development tasks performed using the Java programming language\footnote{\url{http://stackoverflow.com/tags}}. 
In particular, we searched for ``how-to'' development screencasts\footnote{\textit{How-tos} are also among the most requested on Stack Overflow \cite{treude2011programmers}}. 
Therefore, the search string used to retrieve relevant videos from YouTube was   ``Java + \textit{How to}''. 
Our dataset includes 431 Java development screencasts; for all videos a transcript is available. 
We used the Python toolkit pyLDAvis \cite{pyLDA2014, sievert2014ldavis, chuang2012termite} to identify the topics of software development in which the task are performed.    
Using the toolkit, it is possible to visualize different software development topics and to cluster them according to a varying number of LDA topics. 
\begin{figure}[h!]
    \centering
        \includegraphics[width=0.45\textwidth]{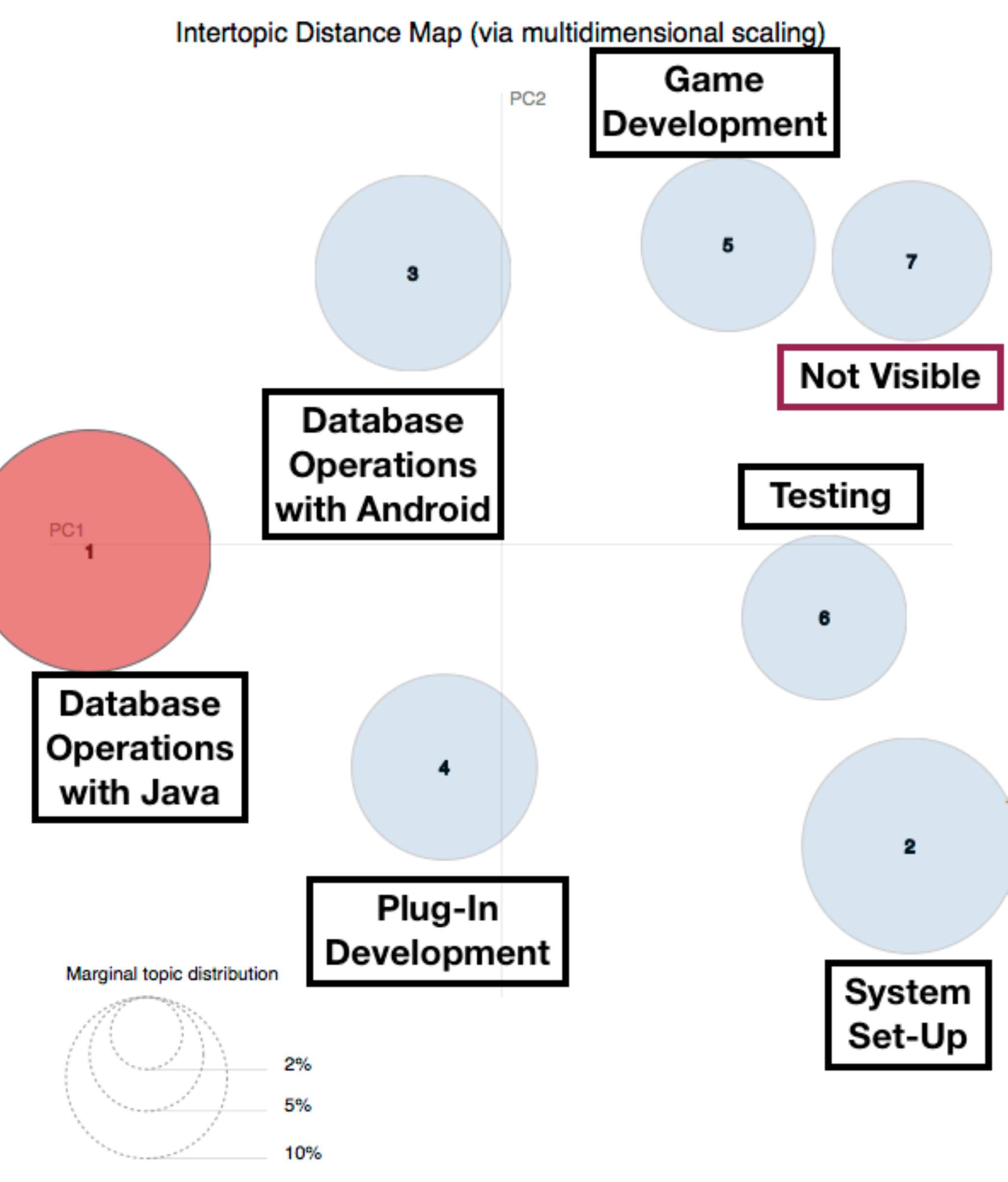}
      \caption{Topics of Java software development topic in relation to other topics.}
      \label{fig:Title_Analysis_Topics}
    \end{figure}

We started by removing from the text all the special characters, numbers and the term ``Java'' which interferes with our analysis.
We tuned the number of LDA topics until we reached a set of non-overlapping clusters that have enough distance between each other (see Figure \ref{fig:Title_Analysis_Topics}).
We also modified the relevance metric $\lambda$ until we found the most relevant terms for a topic of software development        
  \begin{figure}[h!]
    \centering
        \includegraphics[width=0.45\textwidth]{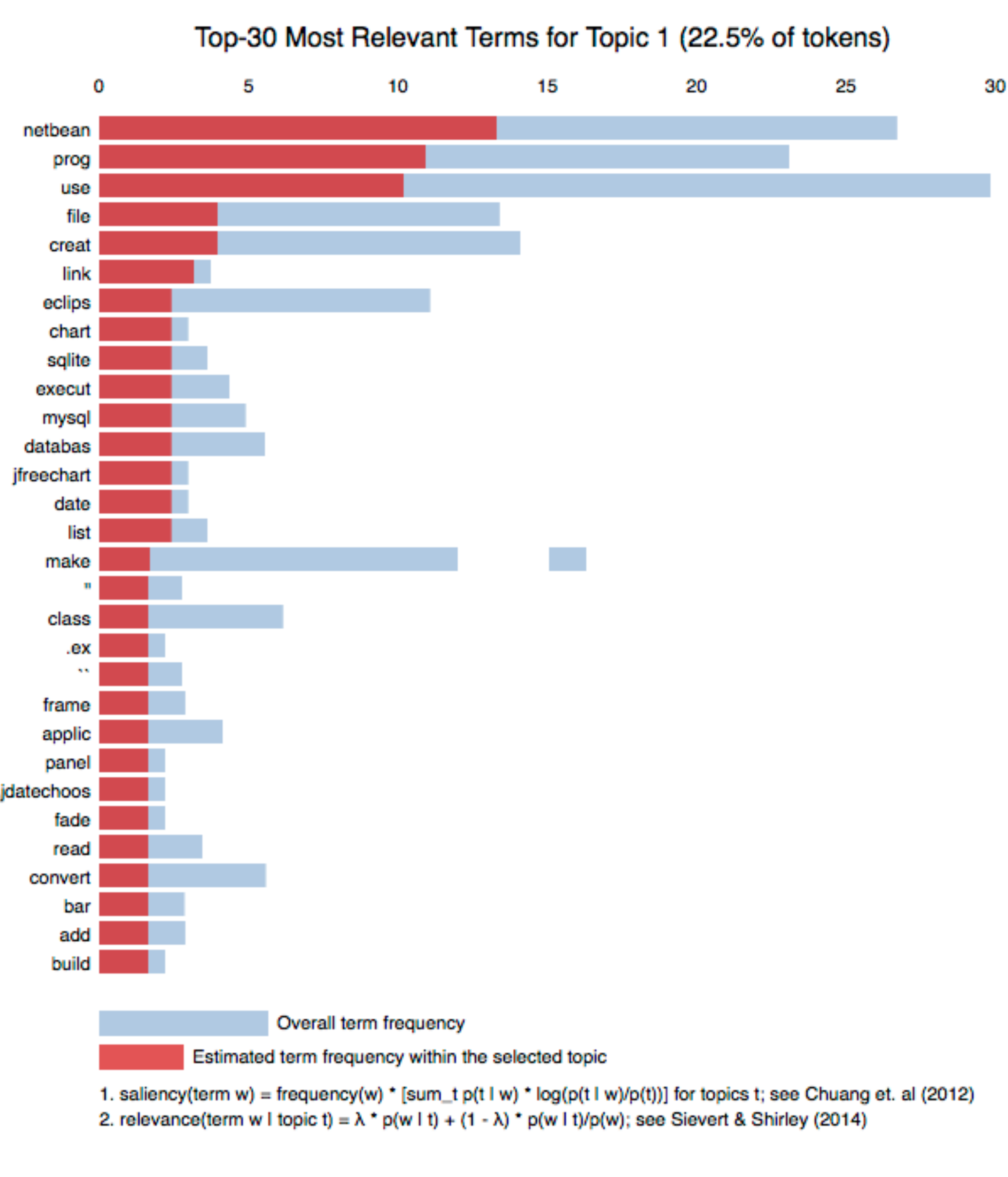}
      \caption{Distribution of terms for a software development screencast topic.}
      \label{fig:Title_Analysis_Tasks}
    \end{figure}
We perform two different analyses of software development topics found in software development screencasts.
In the first analysis, we consider only the titles of the screencast to understand which software development topics is associated with the task performed in the screencast.
The second analysis considers the textual transcript of the development screencasts. 
We only consider the nouns---extracted using the NLTK library \cite{bird2009natural}---since including also verbs caused the algorithm (LDA) to overfit. 
The output of this step was inaccurate because some verbs were included in addition to nouns. 
We listed them in Table \ref{table:topic_screencasts} because we believe that they might be useful for interpreting the overall tasks performed during the screencast. 

Table \ref{table:topic_screencasts} summarizes the topics we found in the titles and in the transcripts of the development screencasts. 
Figure \ref{fig:Title_Analysis_Topics} shows the clusters of all the topics within the chosen software development screencasts. 
We stopped searching for the best number of topics when the topic clusters did not overlap anymore or when the topics became not visible. 
The size of the clusters represents the importance of the topic within the overall set of topics. 
Figure \ref{fig:Title_Analysis_Tasks} shows the distribution of terms used to derive the topic of a task. 

Database-related operations are some of the most popular topics discussed in developers screencasts. 
Similarly, the database management system \textit{MySQL} is one of the most popular topic discussed on StackOverflow\footnote{\url{http://stackoverflow.com/tags}}. 
This observation could indicate the need for a system to support database operations in the IDE. 
Tutorials \cite{tiarks2014does} as well as FAQs \cite{timmann2015} provide a first entry to start developing a certain system. 

Plug-in installation is also discussed in development screencasts. 
This topic can extend traditional tutorials as they provide knowledge for similar development tasks\footnote{\url{http://www.vogella.com/tutorials/EclipsePlugin/article.html}}.
We observed some niche topic, such as game development, discussed in development screencasts. 
Software testing, a frequent software engineering activity\cite{singer2010examination}, is also covered in software development screencasts. 
This might reveal the need for screencasts that teach how to test software \cite{shepard2001more}.
The use of a method, objects, or class in Java is a frequently occurring topic that could be augmented by API reference documents.
In particular, list operations one of the most commonly occurring tasks showing the importance of this data type with respect to similar ones, such as hash-maps. 
Finally, UI operations are also shown to be one of the main activities.

\begin{framed}
\begin{itemize}
\item Database operations are popular development tasks performed in development screencasts. 
\item  Testing---a conventional task in software development--- is also performed in development screencasts.
\item Software development tasks such as methods and classes usage, are taught in software development screencast. 
\item  An advanced search that considers the transcripts of a software development screencast can help finding tasks that matches the development context.
\end{itemize}
\end{framed}

\section{Analysis of similarity to API Documentation}
\label{sec:similarity_analysis}
\begin{figure}[h]
\centering
  \includegraphics[width=0.49\textwidth]{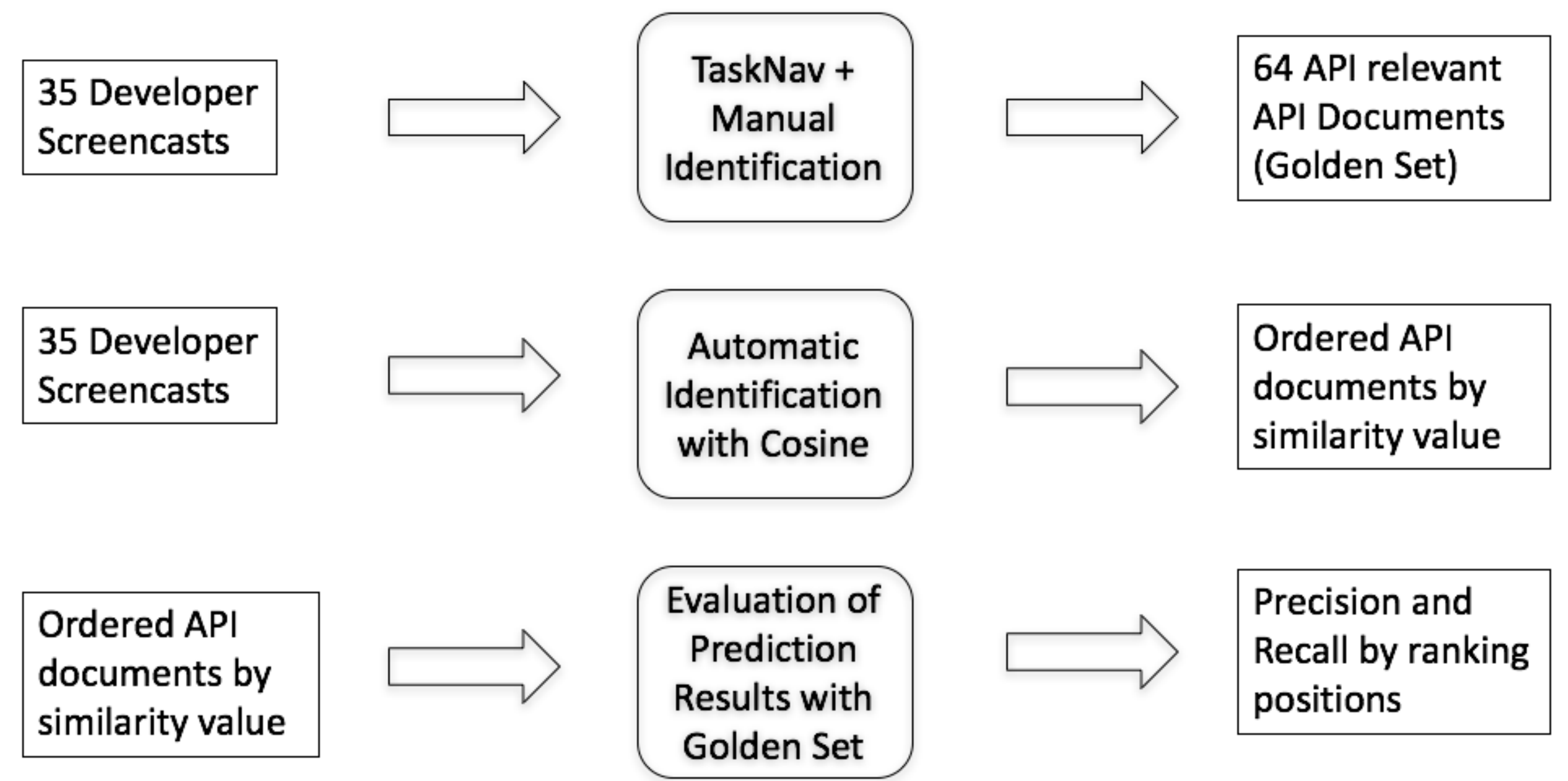} 
  \caption{Method to identify relevant API documents}
  \label{fig:research_method_API_extension}
\end{figure}

\begin{table}[h]
\caption{Prediction results for 65 relevant documents (pages). The search space includes 9,455 API documents.}
\centering

   \begin{center}
       \resizebox{0.85\columnwidth}{!}
               	{

         	  \begin{tabular}{ p{0.3cm} |c|c|c }

         	    \toprule
   Top  & Documents Retrieved  & Precision & Recall\\
   \midrule
   3  &  18/65  & 0.0514 &  0.30\\ \rowcolor{Gray}
   5  &  22/65   & 0.062 &  0.367 \\
   10 &  33/65  & 0.094 & 0.524\\ \rowcolor{Gray}
   20 &  38/65  & 0.0542 & 0.605\\

 	     \bottomrule   
  	  \end{tabular}
  	 }
  	    \end{center}

 \label{table:recall_precsion}
\end{table}

Our dataset contains 35 randomly selected Java development screencasts with high-quality transcripts (e.g., no misspelled words). 
We identified 1-3 relevant API reference documents for every development task that was performed in a development screencast (see research method in Figure \ref{fig:research_method_API_extension}). 
Initially, we used TaskNav---a popular tool for identifying relevant API documents based on natural language\cite{treude2015tasknav}---to find the relevant API document for a development task. 
The input parameter for this tool is a phrase (i.e., the title of the screencast)  describing a certain development task. 
In several occasions, we could not find more than one relevant document because the  screencast titles were not self-explanatory (for example, ``Java How To:  Dialog Boxes'' or ``How to make a Tic Tac Toe game in Java''), and a deeper look into the development screencast and its transcript was often required. 
Therefore, we qualitatively evaluated the recommendations of the API documents by
defining documents as \textit{relevant} if they contain the same classes (e.g., \textit{ArrayList}) or method signatures (e.g., \textit{boolean contains(Object o)}) mentioned in the screencasts as well as additional useful information needed when repeating the development tasks (e.g., \textit{implement ArrayList, LinkedList}).
We could identify 65 relevant documents from 9,455 potential candidates. 
 
For the automatic identification of the relevant development screencasts, we have used the Cosine algorithm. 
We calculated the Cosine similarity value for each transcript of a developer screencast and each of the 9,455 Java API documents in the dataset which resulted in a ranked list of API documents ordered by their similarity values. 
For the evaluation of the recommendations, we calculated precision and recall \cite{robillard2014recommendation} (as identified by TaskNav using manual checking) within the top three, five, 10, and 20 Cosine positions (see Table \ref{table:recall_precsion}).  
Precision shows the percentage of relevant documents identified within a predefined list, whereas recall shows how many relevant documents were identified from all the relevant ones within the same list.

For the best three retrieved results, we found that the transcripts frequently and clearly mention technical terms, such as class and method names contained in an API documentation page. 
Precision varies between 5 and 10\%, with the best result being yielded by the top-10 retrieved pages. 
Table \ref{table:recall_precsion} shows that more than 50\% of the relevant documentation pages were found in the top-10 retrieved positions.
The percentage increases to more than 60 when the top-20 positions are considered.
Overall, we could find 38 out 65 relevant documents until the top-20 in a set of 9,455 potential candidates by just analyzing the screencast transcript and ignoring the text that might appear in a scene (e.g., the source code in the IDE).

Moreover, we found that 98.8\% of the API documents are below a similarity threshold of 0.12 while 55\% of the relevant API documents are above the same threshold.
Considering this threshold when searching for relevant API documents can help developers to find 55\% of the relevant API documents in a list of 114 potential candidates from the overall corpora of 9,444 documents. 
Based on this results, we believe that development screencasts can be extended using API documents considering only their transcript.
\begin{framed}
\begin{itemize}
 
\item By comparing only the audio transcript (the screencast transcripts but not the text that might appear in a scene, e.g. an IDE) of a development screencast with the API documentation, we could identify 38 out of the 65 relevant API documents in the first 20 positions. 
 \item There is a similarity threshold for relevant API documents. A high quantity of relevant API documents can be found above such threshold.
\end{itemize}
\end{framed}

\section{Related Work}
\label{sec:rel_work}
MacLeod et al.\cite{macleod2015code} report on the structure and content of development screencasts, as well as the different types of knowledge located in such screencasts. 
They studied how knowledge was presented and used axial coding extract higher-level topics. 
In our study, we associated Java screencasts to high-level topics, conducted a frame and a similarity analysis, and discussed how screencasts can be used to enrich API reference documentation

Treude et al.\cite{treude2016augmenting} discuss how to link StackOverflow answers to the Java 6 SE SDK API. 
They use the Cosine approach to measure the similarity and LexRank to evaluate the relevance of the API documents. 
We extend their work by linking screencasts with API documents and by showing how similar they are.

Ponzanelli et al.\cite{ponzanelli2016codetube} developed a recommender system to predict relevant YouTube videos\footnote{http://codetube.inf.usi.ch/} for Android development. 
In addition to the audio transcripts, they used an OCR tool\footnote{https://github.com/tesseract-ocr/tesseract/wiki} to transfers the actual video information (e.g., slides  or subtitles) into text. 
They focus on showing relevant StackOverflow posts for random YouTube videos. 

A technique for linking API documentations to code examples is introduced by Subramanian et al.  \cite{subramanian2014live} and  Chen et. al \cite{chen2014asked}. 
They defined a written code as a development task for which an API reference documentation is needed to get insights about the implementation.

\section{Discussion and Limitations}
\label{sec:discussion}
Based on the similarity analysis, we found that frames in a development screencast are much alike in contrast to other types of video. 
Therefore, an identification of screencasts should be possible by using algorithms such as the Cosine similarity or LSI without knowing the actual title, tags, or the transcript of the video. 
Similarly, other types of videos (e.g., recorded interviews or slow motion) are also very static.
We acknowledge that such approach, based on frames comparison, might mistakenly find these other types of \textit{static} videos.

The analysis of the development topics showed that development screencasts contain knowledge provided in API reference documentation. 
Thus, API reference documents can extend a development screencast to provide additional implementation details, making it an attractive media for those developers who do not read documentation \cite{lethbridge2003software}. 
By leveraging our results, a simple tool---e.g., based on Cosine similarity calculation---can suggest relevant documentation pages from a large corpus, like the Java SDK documentation, with a 61\% recall for a list of 20 items.

This preliminary study focuses on screencasts related to a specific programming language. 
However, there is a broad range of other development screencasts which tackle the same topics but in a different manner, or which use different programming languages with different syntax, semantics or specific tools.
Therefore, development screencasts might differ according to the tools used, or to the software engineering activities and phenomena.

The selection of the dataset might thus have influenced the study results.
We used the title to understand the tasks performed in a development screencast, and the transcripts to understand its subtasks. 
Those two elements (i.e., titles and transcripts) complement each other. 
For example, if a developer wants to know how to use lists, files and methods in a programming language like Java she might search them through an algorithm that considers the transcripts.
In this way, the developers can find tasks that match the development context of interest, such as specific IDEs or libraries.

We found that UI operations---one of the most important activity performed when comprehending software \cite{maalej2014comprehension}---are also largely performed in development screencasts. 
By watching screencasts, developers can understand how other developer debugged and solved similar problems. 

The transcripts we obtained might miss important terms, or include misspelled ones. 
This can impact the comparison of those transcripts with the API documentation pages, leading to poor results.
We studied and manually inspected 35 screencasts and their transcripts. 

Building a large dataset using the YouTube API poses some limitations since they only returns a limited number of search results\footnote{http://stackoverflow.com/questions/25918405/youtube-api-v3-page-tokens}. 
Thus, multiple searches, with different search terms, need to be performed. 
Moreover, the persistence of retrieved data is not guaranteed due to the possible deletion of the videos included in our sample.

The  library we used, could not completely identify and remove the verbs or stop words from the title or the transcripts.
Therefore, a replication of this study could lead to different results. 
We recommend to use the NLTK and the pyLDAvis library to pre-process the titles and the transcripts as well as to summarize the topics of the tasks. 
Although different people from different countries might create developments screencasts, we did not evaluate the language quality of the screencasts which might also influence our results.

When performing a development task there is often the need for additional information to be gathered---for example, from Stack Overflow, YouTube or an API documentation. 
Combining all of them mean to use different types of information to perform a development task.

We conclude that software development screencasts can help developers to search for recurring development tasks in a specific context (e.g., within an IDE) independently from the topic of software development.

\section{Conclusion and Future Work}
\label{sec:conclusion}
We analyzed different development screencasts on YouTube and found six main topics for the Java programming language. 
 
A software development screencast is a particular type of video in which developers perform a tasks by focusing on relevant tools.
Development screencasts are not much different from other types of screencasts. 
We found that  frame similarity can be used to detect a development screencast on YouTube.
Development screencasts can be extended by API documents to bette support software developers. 
We found that more than half of the relevant API documents could be provided within a list of 20 items. 
A Cosine comparison between a screencast and a large API documentation corpus is only a preliminary, simple approach to offer developers the most relevant documents. 

This paper provided a first insight on how to categorize and identify development screencasts, and how to enrich them with API documentation. 
A further extension of our approach will focus on extracting the content of the development screencast---e.g., the code showed on the screen when using an IDE---to reach a higher precision/recall when identifying development screencast.

There is also further work needed to determine the different types of knowledge \cite{macleod2015code, Maalej:TSE:2013} located in screencasts to achieve a more fine-grained and precise mapping between the API reference documentation and the API elements within the IDE. 
This approach might require labeling every unique piece of knowledge within a screencast and use video and image features. 
We believe that the community needs to study which types of screencasts are useful for which developers in which situations. 

\bibliographystyle{ieeetr}
\bibliography{bib/FSE2017}

\end{document}